%% file: Allerton09-1.tex
\documentclass[journal,letterpaper,twocolumn]{IEEEtran}
\usepackage{ifthen,subfig,caption}
\newboolean{TwoColumn}
\setboolean{TwoColumn}{true} %
\usepackage{graphicx,verbatim,multirow}
\usepackage{amsmath,amssymb,amsfonts,algorithm2e}
\usepackage{verbatim}
\usepackage[breaklinks=true,letterpaper=true,colorlinks,citecolor=blue,bookmarks=false]{hyperref}
\usepackage[numbers,sort,compress,square,comma]{natbib}
\newcommand{\ignore}[1]{}

\DeclareMathOperator*{\round}{round}
\newcommand{\cut}[1]{{}}
\def\ceil#1{\lceil #1 \rceil}

\newcommand\R{\mathbb{R}}

\newcommand\N{\mathcal{N}}
\newcommand\nbr{\Gamma}
\newcommand\Four{\mathcal{F}}
\newcommand\IFour{\mathcal{F}^{-1}}

\ifthenelse{\boolean{TwoColumn}}{%
 
 \newcommand{\LongEquationStar}[1]{\begin{multline*}#1\end{multline*}}

}{
 
 \newcommand{\LongEquationStar}[1]{\begin{equation*}#1\end{equation*}}
 }

\begin{document}

\author{Danny Bickson, \IEEEmembership{Member, IEEE,}
Dror Baron, \IEEEmembership{Senior Member, IEEE,}
Alexander Ihler, \IEEEmembership{Member, IEEE,}\\
Harel Avissar, \IEEEmembership{Student Member, IEEE,}
Danny Dolev, \IEEEmembership{Member, IEEE}
\thanks{D. Bickson is with the Machine Learning Department, Carnegie Mellon University, Pittsburgh PA 15213, USA. (email: bickson@cs.cmu.edu) Parts of this work were performed when D. Bickson was a researcher at IBM Haifa Research Labs, Israel.}
\thanks{D. Baron is with the Department of Electrical and Computer
Engineering at North Carolina State University, Raleigh, NC 27695,
USA. (email: barondror@ncsu.edu) Parts of this work were performed
when D. Baron was with the Department of Electrical Engineering at the
Technion, Haifa 32000, Israel.
}
\thanks{A. Ihler is with the Bren School of Information and Computer Science, University of California, Irvine CA 92697, USA. (email: ihler@ics.uci.edu)}
\thanks{H. Avissar and D. Dolev are with the School of Computer Science and Engineering, The Hebrew University,   Jerusalem 91904, Israel. (email: harela01@cs.huji.ac.il, dolev@cs.huji.ac.il)}
}%

%

\newcommand{\reals}{{\ensuremath{\mathbb R}}}
\newcommand{\ones}{\mathbf 1}
\newcommand{\diag}{\mathop{\bf diag}}

\newcommand{\conv}{\otimes}
\newcommand{\Conv}{\bigotimes}

\title{ Fault Identification\\ via Non-parametric Belief Propagation}

\markboth{Submitted to IEEE Trans. on Signal Processing}
{Bickson \MakeLowercase{\textit{et al.}}: Fault Identification via
Non-parametric Belief Propagation}

\maketitle

\begin{abstract}
We consider the problem of identifying a pattern of faults
from a set of noisy linear measurements.
Unfortunately,  maximum a posteriori probability  estimation of the
fault pattern is computationally intractable. To solve the fault
identification problem, we propose a non-parametric belief propagation approach.
We show empirically that our belief propagation solver is more accurate than
recent state-of-the-art algorithms including interior point methods and semidefinite
programming. Our superior performance is
explained by the fact that we take into account both the binary nature of the
individual faults and the sparsity of the fault pattern arising from their rarity.
\end{abstract}
\begin{IEEEkeywords}
compressed sensing, fault identification, message passing, non-parametric belief propagation, stochastic approximation.
\end{IEEEkeywords}

\section{Introduction}

\IEEEPARstart{F}{ault} identification is the task of determining which of a set of possible
failures (faults) have occurred in a system given observations of its behavior.
Such problems arise in a variety of applications, including
aerospace \cite{VABDGKMH:07,GDG:04}, industrial process control \cite{HKH:91},
and automotive systems \cite{GCFHZL:93}.

A common approach to fault identification
is to exploit a mathematical model of the system to construct \emph{residuals}
(deviations from the system's expected behavior), which can be used to detect and
diagnose atypical behavior.  A wide variety of approaches have been proposed for
constructing residual measurements.  For example, state space models of the system
may be used to generate parity checks, or the magnitude of errors in a Kalman filter may be used
as an observer to detect changes in behavior.  The occurrence of a particular fault
results in a characteristic change, or signature, which can be used to identify
which fault or set of faults has occurred.
For an overview and survey of residual methods, see Frank~\cite{Frank:90}.

\subsection{Binary faults and sparse signature matrix}

A typical model describes the presence of
each possible fault using a binary variable, and assumes that each fault affects
the observed system measurements in a linear additive way.
Systems of binary faults have
been extensively studied in the computer science community, for example posing the
problem as constraint satisfaction and using heuristic search techniques \cite{Rei:87,dKW:87}.
An alternative approach is to use a convex relaxation of the original combinatorial
problem, for example using the interior point method of
Zymnis et al.~\cite{FaultDet}.

In such systems there are often more potential faults than measurements,
resulting in an under-determined set of linear equations.
By assuming that faults are relatively rare,
one can construct a combinatorial problem to identify the most likely
pattern of faults given the system measurements.


Another common assumption is that the fault signature matrix,
which is the linear transformation matrix applied to the fault vector,
is sparse. This is the case in many
applications, where each fault may affect only a small part (such as a
single subsystem) of the whole~\cite{HJ:97,FSM:93,SDFK:99,BS:85}.
The assumption that the signature matrix is sparse
is especially valid in large scale systems where different faults affect different parts of the system.
Although our algorithm is designed for sparse signature matrices,
we find that it performs well even for moderately dense systems,
as shown in the numerical results section below.

\subsection{Related work}

Fault identification is closely related to several coding and signal reconstruction
problems.  For example, in multi-user detection for code division multiple access (CDMA) systems \cite{GuoWang2008,GuoVerdu2005,bibdb:montanarietal,ISIT2,Allerton},
the system measurements are given by the received noisy wireless signal and the goal is to estimate the
transmitted bit pattern, which plays the role of the fault pattern.
One important difference is that in multi-user detection, each bit
typically has an equal probability of being $0$ or $1$, whereas in fault
identification the prior probability that a bit is $1$ (indicating that a fault is present) is typically much lower.

Compressed sensing (CS) \cite{DonohoCS,CandesCS,CSBP2009,CoSaMP,GPSR,hardIO,GuoBaronShamai2009,RFG2009,limits} is also closely related to fault identification.
Informally, CS reconstructs a sparse signal from a set of (possibly noisy) linear
measurements of the signal.  As in fault identification, the system of linear equations
is ill conditioned, and the assumption of sparsity (corresponding to the rarity of faults)
is critical in the reconstruction.
Because of this similarity between CS and fault identification,
in our numerical results section we compare the performance of several CS
algorithms with previously proposed methods for fault identification.

\subsection{Contributions}

In this work, we develop a novel approach for fault identification based on
a variant of the belief propagation (BP) algorithm called non-parametric belief propagation (NBP).
BP approaches have been applied to solve many similar discrete, combinatorial problems
in coding. NBP  allows the algorithm to reason about real-valued variables. Variants
of NBP have been used in CS~\cite{CSBP2009}
and low-density lattice codes (codes defined over real-valued alphabets)~\cite{LDLC_Sommer,LDLC_Brian,Allerton09-2}.
Our method constructs a relaxation of the
fault pattern prior using a mixture of Gaussians, which takes into
account  both the binary nature of the problem as well as the sparsity of the fault pattern.

Using an experimental study, we show that our approach provides the
best performance in identifying the correct fault patterns when compared to
recent state-of-the-art algorithms, including interior point methods and
semidefinite programming. To demonstrate the importance of each component of
our model, we compare both to existing approaches for fault identification as
well as to CS algorithms and a discrete BP formulation.
We explain how to implement an
efficient quantized version of NBP, and provide the source code
used in our experiments \cite{MatlabGABP}.
Our favorable results can be explained in the context of recent
theoretic results in the related domains
of multi-user detection~\cite{GuoVerdu2005,GuoWang2008}
and CS~\cite{GuoBaronShamai2009,RFG2009,limits}.
In the large system limit, the posteriors estimated by BP converge in
distribution
to the true posteriors, and so NBP is asymptotically optimal.

The structure of this paper is as follows.
Section \ref{s-map} introduces the fault identification problem in terms of
maximum a posteriori (MAP) estimation, and
describes graphical models and the BP algorithm. Section \ref{sec:algo}
presents our solution, based on NBP.
Section~\ref{sec:implement} explains implementation details and optimizations.
Section \ref{sec:exp_results}
compares the accuracy of several state-of-the-art methods for fault
identification, and shows that our proposed method has the highest accuracy.
We shed light on the favorable performance of NBP from an information theoretic
perspective in Section~\ref{sec:inf_theory}, and conclude with a discussion
in Section~\ref{sec:conclusion}.

\section{Fault Identification Problem} \label{s-map}
In this section we describe the fault model in detail, and the basic
maximum a posteriori (MAP)
approach for estimating the fault pattern.
We then briefly review probabilistic graphical models and the belief propagation algorithm
within this context.
Our goal is to infer the MAP fault value,
the fault pattern most likely to have occurred given a set of observations.

\subsection{Fault model and prior distribution}
We consider a system in which there are $n$ potential faults, any
combination of which ($2^n$ in total) can occur.
A fault pattern, i.e., a set of faults, is represented by a vector
$x \in \{0,1\}^n$, where $x_s=1$ means that fault $s$\,, $s\in \{1,\cdots,n\}$ has occurred.
We assume that faults are independent and identically distributed (i.i.d.), and that
fault $s$ occurs with known probability $p_s$.
Thus, the (prior) probability of fault pattern $x$ occurring is
\begin{equation*}
p(x) = \prod_{s=1}^n p_s^{x_s}\left(1-p_s\right)^{1-x_s}.
\end{equation*}
The fault pattern $x=0$ corresponds to the null hypothesis, the
situation in which no faults have occurred, with
probability $p(0)=\prod_{s=1}^n (1-p_s)$.
The expected number of faults is $\sum_{s=1}^n p_s$.

We assume that $m$ scalar real measurements, denoted
by $y$, $y \in \reals^m$, are available.  These measurements
depend on the fault pattern $x\in \{0,1\}^n$ linearly:
\begin{equation*}
y=Ax+v\,,
\end{equation*}
where $A \in \reals^{m \times n}$ is the \emph{fault signature matrix},  and
the measurement noise $v \in \reals^m$ is random, with
$v_{i}$ independent of each other and $x$,
each with $\N(x;0,\sigma^2)$ distribution.
Typically the system of linear equations is under-determined ($n > m$),
which means the number of potential faults is greater than the number of measurements.

The fault signature matrix $A$ is assumed to be known.
Its $s^{th}$ column $a_s \in \reals^m$ corresponds to the measurements,
when only fault $s$ has occurred and assuming no noise.
For this reason $a_s$ is called the $s^{th}$ \emph{fault signature}.
We further denote $\tilde{a}_j \in \R^n$ as the $j^{th}$ row of the matrix $A$.
Since $x$ is a Boolean vector, $Ax$ is the sum
of the fault signatures corresponding to the faults that have occurred.
We further assume $A$ to be sparse, i.e.,
the percentage $q$ of the non-zero values of $A$ is much smaller than one.
Indeed, the matrix $A$ is sparse in many
applications, where each fault may affect only a small part (such as a
single subsystem) of the whole~\cite{HJ:97,FSM:93,SDFK:99,BS:85}.

\subsection{Analog circuit example}
\label{analogc}
As an example, consider
linear analog circuits, in which it is common to use \emph{nodal
analysis} to describe the behavior of the
system; non-linear circuits may be approximately linearized to apply a
similar approach.
Each internal node of the circuit is
assigned a variable representing its voltage $v_i$, and voltage and
current sources are given variables indicating
their induced current or voltage, respectively.
Applying Kirchoff's current law, we know that the total current flowing
from any internal node $i$ must be zero; when these currents are
written in terms of the nodal voltages they yield
a set of linear equations $Av=w$ that describe the system,
where $w_i=0$ for internal nodes, and equals the known voltage/current
of a fixed voltage/current source.

For example, for a sequence of three nodes $i$, $j$, $k$ connected
to their neighbors with resistances $R_{ij}$, $R_{jk}$, Kirchoff's law
applied to $v_j$ yields
$\frac{v_j-v_i}{R_{ij}} + \frac{v_j-v_k}{R_{jk}}=0$.  One can solve
for the internal voltages easily as
$v=A^{-1}w$; we assume that some subset of these voltages are
measurable for testing, $v_M=MA^{-1}w$, where $M$ is a measurement matrix.

A fault, including an incorrect component value, short, or
open-circuit, then corresponds to a localized change
in the matrix $A$ associated with those two nodes.  For example, a
change $R_{jk}\rightarrow R'_{jk}$ adds the term $(v_j-v_k)(\frac{1}{R'_{jk}}-\frac{1}{R_{jk}})$ to row $j$ of $A$, and
subtracts it from row $k$.
If the resulting new, faulty circuit matrix is $A'$,
the fault signature is then $v'_M-v_M = M(A'^{-1}-A^{-1})w$.
Note that the underlying assumption is that the number of actual faults in a single circuit is small, which explains the imbalance in the fault prior probability $p << 1$. For more information about detecting faults in linear analog circuits, see e.g.~\cite{AnalogFaultsBook}. 
\subsection{Posterior probability}
\label{subsec:posterior_prob}

Let $p(x|y)$ denote the (posterior) probability of pattern $x$ given
the measurements $y$.  By Bayes rule we have
\begin{align}
p(x|y) &\propto p(y|x) p(x) = p(x,y) \notag \\
& = \N(y\ ;\ Ax, \sigma^2 I) \prod_s p_s^{x_s}(1-p_s)^{1-x_s}\,. \label{probfunc}
\end{align}
Letting $C$ and $C'$ indicate constants (values independent of $x$), we can
define the log-loss function as the negative log probability
$l_y(x) = -\log p(x|y)$ and write
\begin{align}
l_y(x) &= \lambda^Tx +\tfrac{1}{2\sigma^2} \sum_{i=1}^m (y-Ax)^T (y-Ax)+ C \notag \\
       &= \tfrac{1}{2\sigma^2}x^TA^TAx + (\lambda -\sigma^{-2}A^Ty)^Tx + C'\,, \label{costfunc}
\end{align}
where $\lambda_s = \log((1-p_s)/p_s)$ denotes the log-odds ratio.
Note that~\eqref{costfunc} is a convex quadratic function of $x$
(a binary-valued vector),
and MAP estimation is thus a convex integer quadratic program.

\subsection{Graphical Models}
\label{sec:graphical_models}
Graphical models are used to represent and exploit the \emph{structure} of the cost
function~\eqref{costfunc}, or its associated probability distribution~\eqref{probfunc},
to develop efficient estimation algorithms.
Specifically, we factor $p(x|y)$ into a product of smaller
functions, called factors, each of which is defined using only a few variables.
This collection of smaller functions is then represented
as a factor graph $G$, in which each variable is associated with a \emph{variable node}
and each factor with a \emph{factor node}.  Factor nodes are connected to variable
nodes that represent their arguments.
Because~\eqref{costfunc} can be represented as the sum of terms involving
at most two variables,
\begin{equation} \label{eq:gm1}
l_y(x) = \sum_{s,t>s} J_{st} x_s x_t + \sum_s h_s x_s\,,
\end{equation}
where
\begin{align*}
J_{st} &=  \tfrac{1}{\sigma^2}  (A^T A)_{st}\,, \\
h_s &= (\lambda- \tfrac{1}{\sigma^2} A^T y)_s + \tfrac{1}{2\sigma^2} (A^T A)_{ss}\,,
\end{align*}
and we can represent $p(x|y)$ or $l_y(x)$ as a pairwise graph with an edge between $x_s$
and $x_t$ if and only if $J_{st}$ is non-zero.  This corresponds to factoring $p(x|y)$ as
\begin{align*}
p(x|y) &\propto \prod_{i=(s,t>s)} f_{i}(x_s,x_t) \prod_s g_s(x_s) \\
 &= \prod_{(s,t>s)} \exp(-J_{st} x_s x_t) \prod_s \exp(-h_s x_s)\,,
\end{align*}
where $f_i(\cdot)$ and $g_s(\cdot)$ are factors; we use the convention that
functions $g$ and indices $s,t$ refer to local (single-variable) factors while $f$
and indices $i,j$ refer to higher-order (in this case, pairwise) factors.  For compactness,
we generically write $f(x)$ to indicate a function $f$ over some subset of the $x_s$.

The graph structure is used to define efficient inference algorithms, including
MAP estimation or marginalization.
Both problems are potentially
difficult, as they require optimizing or summing over a large space of possible
configurations.  However, structure in the graph may induce an efficient method of
performing these operations.  For example, in sequential problems the Viterbi algorithm~\cite{Viterbi}
(and more generally, dynamic programming)
provides efficient optimization,
and can be represented as a message-passing algorithm on the graph $G$.

In graphs with more complicated structure such as cycles, exact inference is often
difficult; however, similar message-passing algorithms such as loopy belief propagation
perform approximate inference \cite{NBP}.  The max-product algorithm is a form
of BP that generalizes dynamic programming to an approximate algorithm.
One computes messages $m^f_{is}$ from factors to variables and $m^v_{si}$ from variables
to factors,
\begin{eqnarray} 
m^f_{is}(x_s) \propto \max_{x\setminus x_s} f_i(x) \prod_{t\in\nbr_i\setminus s}m^v_{ti}(x_t)\,,\label{eq:maxprodA} \\
m^v_{si}(x_s)     \propto \ \ \ \ g_s(x_s) \prod_{j\in\nbr_s\setminus i}m^f_{js}(x_s)\,,
\label{eq:maxprodB}
\end{eqnarray}
where messages are normalized for numerical stability, and $\nbr_s$ is the neighborhood
of node~$s$ in the graph (all nodes for which $\tilde a_i$ is non-zero, excluding $s$).
One can also compute a ``belief'' $b_s(x_s)$ about variable $x_s$,
\begin{equation}
b_s(x_s) = g_s(x_s) \prod_{i\in\nbr_s} m^f_{is}(x_s)\,,
\end{equation}
which can be used to select the configuration of $x_s$ by choosing its maximizing value.
If the graph $G$ is singly-connected (no cycles), then the max-product algorithm is
equivalent to dynamic programming.  However, the algorithm performs well even in graphs with
cycles, and has been shown to be highly successful in many problems, most notably
the decoding of low-density parity check (LDPC) codes~\cite{BibDB:Gallager}.  Max-product
and its so-called reweighted variants are closely related to linear programming relaxation
techniques \cite{TreeReweighted,TreeReweighted2,wainwright05a}, but by exploiting the problem structure can be more efficient
than generic linear programming packages~\cite{yanover06}.

The sum-product formulation of BP is intended for approximate
marginalization, rather than optimization.  Despite this, the sum-product algorithm
has been frequently
applied to MAP estimation problems, as it often exhibits better convergence behavior
than max-product \cite{SS}.  It has an almost identical message-passing formulation,
\begin{eqnarray} 
m^f_{is}(x_s) &\propto \sum_{x\setminus x_s} f_i(x) \prod_{t\in\nbr_i\setminus s}m^v_{ti}(x_t)\,,\label{eq:sumprod} 
\end{eqnarray}
where the product step is computed as in \eqref{eq:maxprodB}. Again, one can estimate each $x_s$ by choosing
the value that maximizes the belief $b_s(x_s)$, in this case corresponding to
the maximum posterior marginal estimator.  Although the sum-product
beliefs are intended to approximate the marginal distributions of each $x_s$,
if the most likely joint configurations all share a particular value for
$x_s$, then this will be reflected in the marginal probability as well.
Finally, a connection between non-asymptotic block length sum-product belief propagation and joint maximum likelihood or MAP\ detection was described in \cite{walsh10}.

Variants of BP typically
rely on graph sparsity (few edges) to ensure both efficiency and accuracy.
When the graph has no cycles, these algorithms are exact; for graphs
with cycles, they are typically only approximate but are often
accurate in systems with long, weak, or irregular cycles~\cite{ihler07b}.
Unfortunately,  although the fault signature matrix $A$ may be sparse,
the same is typically not true of the matrix $J=A^T A$, especially for large $m$ and $n$.
In the dense case, a direct application of BP to~\eqref{eq:gm1} may fail.
For example, in experiments with $A$ sized $50\times 100$ and approximately $10\%$ non-zero values $\pm 1$,
$A^T A$ is approximately $50\%$ non-zero,
and BP algorithms defined using~\eqref{eq:gm1}
did less well than the current state of the art methods.
As the dimension sizes $m$ and $n$ are increased, $A^T A$ becomes dominated by non-zeros.
This motivates us to define an alternative graphical model that
relies on the sparsity of $A$ itself.
%

%


\section{Belief Propagation for Fault Identification}
\label{sec:algo}
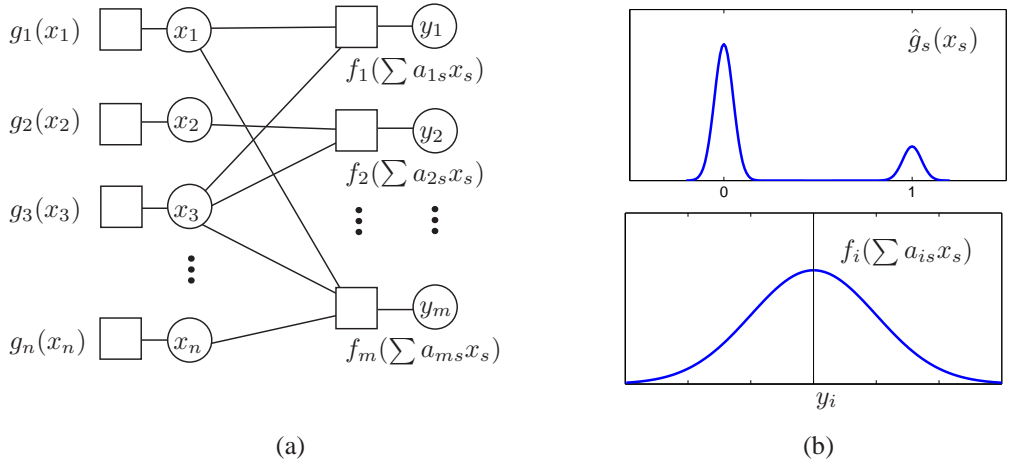
\begin{figure*}[t] \centering
\begin{tabular}{cc}
 \hspace{1cm}
 \includegraphics[width=2in]{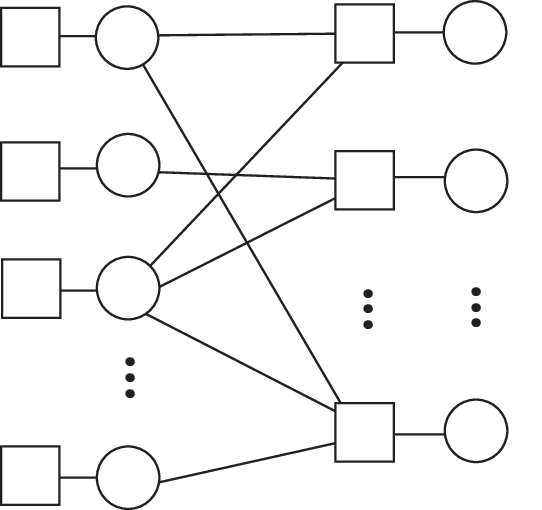}
\begin{picture}(0,0)
\put(-182,123){$ g_1(x_1)$}
\put(-120,123){$x_1$}
\put(-182,89){$ g_2(x_2)$}
\put(-120,89){$x_2$}
\put(-182,55){$ g_3(x_3)$}
\put(-120,55){$x_3$}
\put(-182,5){$ g_n(x_n)$}
\put(-120,5){$x_n$}
\put(-55,108){$f_1(\sum  a_{1s} x_s)$}
\put(-27,125){$y_1$}
\put(-55,69){$f_2(\sum  a_{2s} x_s)$}
\put(-27,85){$y_2$}
\put(-55,2){$f_m(\sum  a_{ms} x_s)$}
\put(-27,19){$y_m$}
\end{picture}\hspace{1cm}
&
\raisebox{.8in}{
\begin{tabular}{c}
\includegraphics[width=2in]{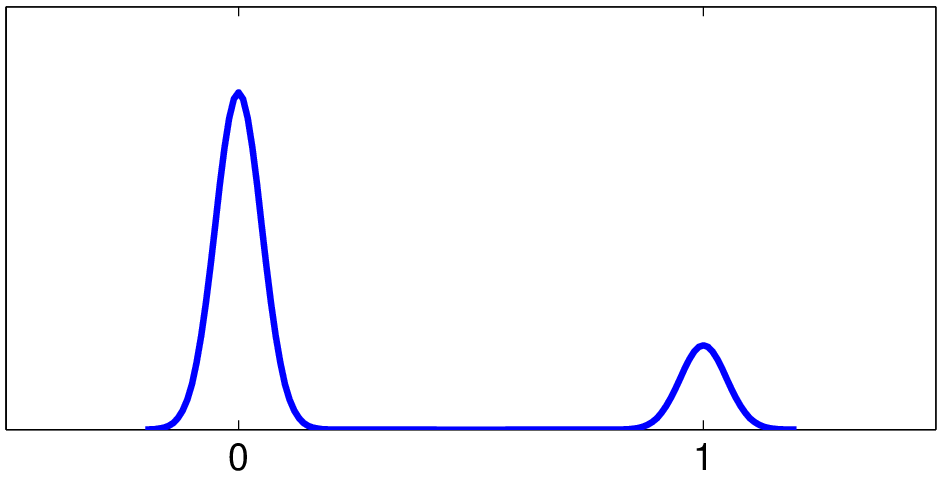}\\
\includegraphics[width=2in]{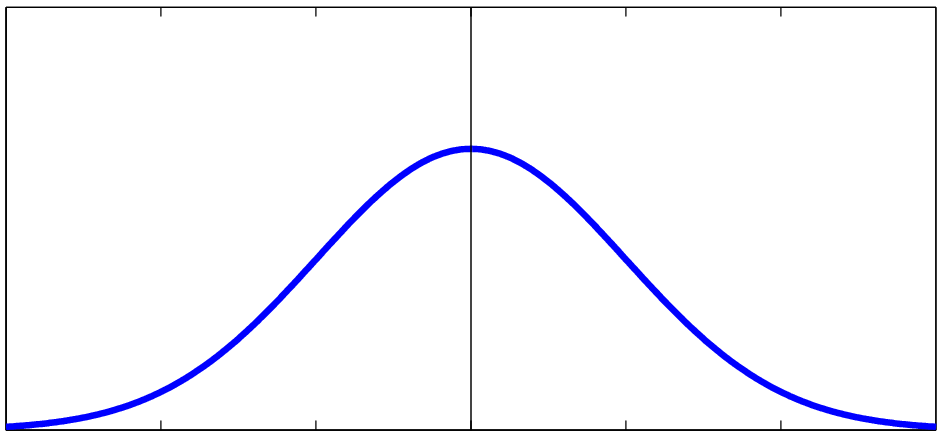}
\begin{picture}(0,0)
\put(-40,130){$\hat g_s(x_s)$}
\put(-66,50){$f_i(\sum  a_{is} x_s)$}
\put(-75,-5){$y_i$}
\end{picture}
\end{tabular}
} \\
\\
(a) & (b)
\end{tabular}
\caption{Graphical model for the fault identification problem.
(a) Factor graphs represent dependence among variables,
including faults $\{x_s\}$ and measurements $\{y_i\}$.
Graph edges indicate non-zero values of $\tilde a_{i}$.
(b) Potential functions $\hat g_s$ capture the binary and sparse nature of the $x_s$, and $f_i$
represent the linear measurements with Gaussian noise.
}
\label{fig:factorgraph}
\end{figure*}

Since the structure of the matrix $A$ is sparse, let us define an alternative graphical
model that uses $A$ explicitly.  In particular, we can write the probability distribution
\[ p(x,y)=\prod \limits_i f_i(y_i,x) \prod \limits_s g_s(x_s)\,,\] in terms of the factors,
\begin{equation}
f_i(y_i,x) = \N(y_i;\tilde a_i x,\, \sigma^2), \quad
g_s(x_s) =  p_s^{x_s} (1-p_s)^{1-x_s}.
\label{eq:ctsgraph}
\end{equation}
Notably, if $A$ is sparse (specifically, if each row $\tilde a_i$ is sparse), then the
factors $f_i$ will depend only on the few $x_s$ for which $\tilde a_i$ is non-zero
and the graph representing this factorization will be sparse as well.  An example
factor graph is shown in Fig.~\ref{fig:factorgraph}(a).

Using either max-product~\eqref{eq:maxprodA} or sum-product~\eqref{eq:sumprod} on the 
resulting factor graph is computationally difficult, as it involves eliminating (maximizing
or summing over) all exponentially many configurations of the neighboring variables of $f_i$
($2^d$ evaluations for factors over $d$ variables).  This can quickly become intractable for
even moderate neighborhood sizes.  Although the relationships among the $x_s$ and $y_i$
are simple and linear, our model defines a hybrid distribution over both continuous valued
and discrete valued random variables.  Observing $y_i$ creates a combinatorial dependence
among the neighboring discrete-valued $x_s$.  Again, in our experiments with $n=100$, $m=50$,
these computations became extremely slow for even 
modest values of $q$;
for example, when $q\approx 0.15$, even the average factor has $d=15$ and the largest factor is
typically significantly higher.


Although it may seem counter-intuitive, we can avoid some of these difficulties by
converting the graphical model to a fully continuous model.
We abuse notation slightly to define continuous
variables $x_s$ in place of their discrete counterparts, with corresponding factors
\begin{align*}
g_s(x_s) &= p_s \delta(x_s = 1) + (1-p_s) \delta(x_s=0)\,.
\end{align*}
It will also prove convenient to relax the prior slightly into a Gaussian form,
using the approximation
\begin{align}
\hat g_s(x_s) &= p_s \N(x_s;1,\nu) + (1-p_s) \N(x_s;0,\nu)\,, \label{gauss_approx}
\end{align}
where the variance $\nu$ controls the quality of the approximation; as $\nu \rightarrow 0$,
we recover the original prior on $x_s$.
The factors $f_i(\cdot)$ and $g_s(\cdot)$ are illustrated in Fig.~\ref{fig:factorgraph}(b).

The BP message-passing algorithm remains applicable to models defined
over continuous random variables.  However, message representation is more difficult as we
must represent continuous functions over those variables. For non-Gaussian models,
message representation typically requires some form of approximation.

In our continuous model, both sets of factors $f_i$
and $\hat g_s$ consist of mixtures of Gaussians; thus their product, and the messages
computed during BP, will also be representable as mixtures of
Gaussians~\cite{NBP,LDLC_Sommer}.
Unfortunately, at each step of the algorithm, the number of mixture components required to
represent the messages will increase at an exponential rate, and must be approximated
by a smaller mixture.
To handle the exponential growth of mixture components,  approximations to BP over continuous variables
were proposed in NBP~\cite{NBP}
and variants  in a broad array of problem domains~\cite{Ihler-JSAC,LDLC_Sommer,LDLC_Brian,CSBP2009}.
Those approximations use sampling to limit the number of components in each mixture.
A number of sampling algorithms have been designed to ensure that  sampling  is
efficient~\cite{NBP2,briers05,rudoy07}. Such sample based representations
are particularly useful in high dimensional spaces, where discretization becomes computationally
difficult.
Various authors have also proposed message approximation methods based on
dynamic quantization or discretization techniques~\cite{coughlan07,isard09}.

For the fault identification problem our variables $x_s$ are one dimensional and can
be reasonably restricted to a finite domain (Section \ref{boundx}).
Thus, it is both computationally efficient and sufficiently accurate to use a
simple uniform discretization over possible values, allowing our functions over $x_s$ to
be represented by fixed-length vectors~\cite{CSBP2009,LDLC_Sommer}.
Although typically the term ``non-parametric BP'' refers to an algorithm
using stochastic samples and Gaussian mixture approximations to the messages, rather than a fixed discrete
quantization, here we use it more generically to distinguish BP in our
fully continuous model from a standard discrete BP performed
directly on~\eqref{eq:gm1}.

\section{Implementation}
\label{sec:implement}
In this section we discuss some details of our implementation of NBP
and our overall fault identification algorithm.  These include several techniques
that make our algorithm more efficient, and two local optimization
heuristics that may improve solution quality.

\subsection{Computation of messages} \label{subsec:messages}
We apply the BP algorithm presented in \eqref{eq:sumprod} using
the factors \eqref{eq:ctsgraph} and the self potentials
defined by the Gaussian approximation \eqref{gauss_approx}.
The variable-to-factor messages $m^v_{si}(x_s)$ are given 
as in~\eqref{eq:sumprod},
and we compute the message product more efficiently by noting that
\begin{equation}\label{eq:divide}
g_s(x_s) \prod_{j\in\nbr_s\setminus i}m^f_{js}(x_s) = \frac{g_s(x_s) \prod_{j\in\nbr_s} m^f_{js}(x_s)}{m^f_{is}(x_s)}\,.
\end{equation}
The product on the r.h.s of \eqref{eq:divide} can be computed once and reused for each outgoing
message~\cite{BroadcastBP,Ihler-JSAC}.

The factor-to-variable messages require a more detailed analysis.
It is important to note that our factors $f_i(x)$ are functions of
several variables, say $\nbr_i=\{x_{s_1}\ldots x_{s_d}\}$. A message computation,
for example from factor $i$ to variable $s_1$, can be explicitly written as
\begin{align}\label{eq:explicit}
  m^f_{is_1} \propto \int_{x_{s_2}}\!\!\!\! \ldots\int_{x_{s_d}} \!\!\!\! \Big[f_i\big(\sum_{j=1}^d  a_{is_j}x_{s_j}\big) \prod_{j=2}^d m^v_{s_{j}i}(x_i)\Big]dx_{\Gamma_{i}}\,.
\end{align}
Although an arbitrary function over $d$ variables would require $O(n^d)$ computation, where $n$
is the number of discretization bins, the messages can be computed more
efficiently, because $f_i$ is a function over a linear combination of the $x_s$ \cite{potetz08,LDLC_Sommer}.
In essence, one uses a change of variables to separate
each integrand, defining variables for the cumulative sum,
$\bar x_{s_j} = a_{is_j} x_{s_j} + \bar x_{s_{j+1}}$, and scaled messages
$\bar m_{s_ji}(x) =  m_{s_ji}(x/ a_{is_j})$. We re-express~\eqref{eq:explicit},
\LongEquationStar{
  m^f_{is_1}(x_{s_1}) \propto \\
  \int_{\bar x_{s_2}} \!\!\!\! f_i( a_{is_1}x_{s_1}+\bar x_{s_2})
  \int_{\bar x_{s_3}} \!\!\!\! \bar m_{s_2i}(\bar x_{s_2}-\bar x_{s_3}) \int_{\bar x_{s_4}} \!\!\!\!\!\ldots \bar m_{s_di}(\bar x_{s_d})d\bar{x}_{\Gamma_{i}}, 
}
in which each integral (approximated by a discrete sum) requires $O(n^2)$ computation.
Note also that, computing from the right-hand side, each step takes the form
$m_j(\bar x_j)=\int \bar m(\bar x_j - \bar x_{j+1})m_{j+1}(\bar x_{j+1})dx_{j+1}$,
and can thus be thought of as a convolution operator, $m(\bar x_j)\conv m(\bar x_{j+1})$~\cite{felzenszwalb06}.
For convenience, we write this convolution as
\begin{align} \label{eq:FtoV}
  m^f_{is}(x_s) \propto f_i(a_{is}x_s) \conv \Conv_{t\in \nbr_i\setminus s} m^v_{ti}(x_t/a_{it})\,.
\end{align}
Moreover, since convolution can be computed by an element-wise product in the Fourier domain,
the factor-to-variable messages can be evaluated even more efficiently by first rescaling each incoming message,
transforming into the Fourier domain, taking a product, transforming back, and unscaling,
resulting in the update rule~\cite{felzenszwalb06},
\begin{align} \label{eq:Fdomain}
  m^f_{is}(x_s/a_{is}) \propto \IFour\bigg( \Four(f_i) \prod_{t\in \nbr_i\setminus s} \Four(m^v_{ti}(x_t/a_{it})) \bigg)\,,
\end{align}
where $\Four$ and $\IFour$ are the discrete Fourier and inverse Fourier transforms,
respectively. Again, the products are computed more efficiently using all terms and dividing as in~\eqref{eq:divide}.

This highlights the basic advantage over a formulation in which the $x_s$ are explicitly discrete-valued.
Although the exact calculations are exponential over the degree of the factor $f_i$ in both cases,
the continuous-valued formulation provides us the opportunity to approximate the intermediate quantities
(in our implementation, using a discretization) and separates the computation into a simple and
efficiently computed form~\eqref{eq:Fdomain}.
We note that the rescaling step is equivalent to the stretch/unstretch operations
proposed in LDLC \cite{LDLC_Sommer}.
Additionally, if the scale factors $a_{it}$ are bipolar ($\pm 1$), then rescaling becomes trivial
(e.g., for $-1$, the vector is reversed), and the algorithm simplifies further.

We proceed by computing all messages from variables to factors according
to~\eqref{eq:divide}, then computing all messages from factors to variables
\eqref{eq:Fdomain}.  The algorithm is run for a predetermined number of
iterations, or until convergence is detected locally. To detect convergence, we use the $\ell_2$
norm of the product of all incoming messages in the current iteration, relative to
the product of all incoming messages in the previous iteration.  Finally, each
variable node $x_s$ computes its belief, and we estimate its value by
rounding to the closest fault value (either zero or one),
\begin{equation} \label{eq:finally}
x_s = \round\left( \arg \max_{x_s}\left\{g_s(x_s) \prod_{i\in \nbr_s} m^f_{is}(x_s)\right\} \right)\,.
\end{equation}

\subsection{Discretization of the Gaussian mixture}
\label{boundx}
Recall that messages are Gaussian mixtures representing the posterior and are discretized for efficiency. To
store each message, we allocate a vector of  fixed length $b$. Entries in this vector
are real positive values. Typically we use values
of $b$  in the range $512-1024$.  A higher value of $b$ makes the algorithm more accurate
but slows execution time. We evaluate the messages at
fixed intervals $\Delta$ within the range of interest, identical for each variable.
This range should include, for example, both 0 and 1 when the fault pattern is binary
and the actual real-valued measurements; because of noise,
the range of discretization is further increased to include several standard
deviations of uncertainty.
We used the following heuristic formula to determine the scope of discretization:
\begin{equation}
R = 1.2\max_i \max\{|y_i|,\sum_s |a_{is}|+3\sigma\}\,,\label{eq:bins}
\end{equation}
and centered around zero, i.e., the range $[-R,R]$.
For Bernoulli matrices $A$, $|a_{is}|=1$ and when $qn \gg \sigma$, 
we can further simplify the right-hand term to $qn$.
This range ensures that the range is great enough to include the partial sums of each message, plus
some noise;
the symmetry condition is useful for computing the scaling
operation on negative-valued edges, by first computing the scale then reversing the output around zero.

Note that multiplication of two FFTs followed by an inverse FFT \eqref{eq:Fdomain} is
not equivalent to linear convolution, but rather circular convolution, and therefore zero
padding must be carried out. We selected the bins for quantization \eqref{eq:bins} to be
large enough such that no significant probability is near the edges, and so any
aliasing is minimal.

\subsection{Local optimization procedures}
\label{s-heuristics}
We outline two useful heuristics proposed by Zymnis et al.~\cite{FaultDet} that
are used for improving the quality of our solutions: variable threshold rounding
and a local search procedure. The purpose of these
heuristics is to obtain an integer solution out of the fractional solution obtained using our BP solver.

\ifthenelse{\boolean{TwoColumn}}{
  \input{table1.inc} 
}{}
\newcommand{\xsrmap}{x^+}

\paragraph{Variable threshold rounding}
Let $x^\star$ denote a \emph{soft decision}, a vector of the same length as $x$
whose entries are real-valued (rather than binary); $x^\star$ may correspond,
for example, to the BP belief or to the optimal point of a convex relaxation
of the original problem.
Our task is to round the soft decision $x^\star$
to obtain a valid Boolean fault pattern (or \emph{hard decision}).
Let $\theta \in (0,1)$ and set
\begin{equation*}
\hat x = \ceil{(x^\star-\theta)}.
\end{equation*}
To create $\hat x$, we simply round all entries of $x^\star$ smaller than
the threshold $\theta$ to zero.  Thus $\theta$ is a threshold for
guessing that a fault has occurred, based on the
relaxed MAP solution $x^\star$.
As $\theta$ varies from $0$ to $1$, this method generates up to $n$
different estimates $\hat x$, as each entry in $x^\star$ falls below
the threshold.
We can efficiently find them all by
sorting the entries of $x^\star$, and setting the
values of each $\hat x_s$ to one in the order of increasing $x^\star_s$.
We evaluate the loss for each of these patterns \eqref{costfunc}, and take as our best 
fault estimate the one that has least loss, which we denote by $\xsrmap$.

\paragraph{Local search}
Further improvement of the estimate can  be obtained by performing
a local search around $\xsrmap$. Initializing $\hat x$ to $\xsrmap$,
we cycle through indices $s=1,\ldots, n$, where at step $s$ we replace $\hat x_s$
with $1-\hat x_s$.
When this change decreases the loss function, the change is accepted; we then move on to the next index.
We continue in this manner until we have rejected changes in all entries of $\hat x$.
At the end of this search, $\hat x$ is at least 1-OPT, which means
that no change in any one entry will improve the loss function.

\ifthenelse{\boolean{TwoColumn}}{}{
  \input{table1.inc} 
}
\section{Numerical Results}
\label{sec:exp_results}
\subsection{Algorithms  compared}
We have implemented our NBP solver using Matlab; our implementation is available
online~\cite{MatlabGABP}. Table~\ref{tab:methods} lists the different algorithms we evaluated.
We compared NBP to several groups of competing state-of-art algorithms.
First, we considered the interior point method (IP) for solving
the fault identification problem~\cite{FaultDet}.
Second, we evaluated two other variants of NBP:
({\em i}) CSBP~\cite{CSBP2009}; 
and ({\em ii}) the low density lattice decoder (LDLC)~\cite{LDLC_Sommer}.
Third, we ran several non-Bayesian CS algorithms:
({\em i}) CoSaMP \cite{CoSaMP};
({\em ii}) GPSR \cite{GPSR}; and
({\em iii}) iterative hard thresholding (HardIO) \cite{hardIO}.
Fourth, we implemented a semidefinite programming relaxation~\cite{TR:01,ANJM:02}.
Finally, for a MAP problem over binary variables, it is natural to employ
the discrete BP max-product algorithm defined directly over the model~\eqref{eq:gm1}, and so we implemented
this algorithm as well. In practice, max-product BP performed significantly worse than  other algorithms.
We also evaluated a factor graph representation with binary-valued variables, implemented in
C++ using the libDAI toolbox~\cite{Mooij_libDAI_10}.
For small values of $q$ the factor graph BP was as effective as NBP, but
as discussed in Section~\ref{sec:algo} it requires time exponential in the size of the largest
factor; for $q=0.15$ it required on average 3 minutes per execution, compared to 1.6 seconds for NBP.


A technical subtlety that arises when handling the various algorithms is that
they use one of two equivalent formulations of the problem: either a
Boolean or bipolar representation. Table~\ref{tab:xform} outlines the two
models and the transformation needed to shift between them; we use the
notation $\ones$ for the all-ones vector of appropriate size. Without loss
of generality we use the binary form when describing the algorithms.

\begin{table}[t]
\centering
\begin{tabular}{|c|c|c|}\hline
Bipolar & Binary &\ Transformation \\ \hline \hline
$ x \in \{-1,1\}^n$ & $\overline{x} \in \{0,1\}^n$ & $ \overline{x} = (x+1)/2$\\
$ y = Ax+v$ & $ \overline{y} =(2A)\overline{x} + v $ & $\overline{y} = y+A\ones$\\
$ \min \limits_x \|Ax-y\|  $ & $ \min \limits_{\overline{x}} \|(2A)\overline{x}-\overline{y}\|$ &  \\
$ \quad \mbox{s.t.}\ x \in \{-1,1\}^n$ & $\quad \mbox{s.t.}\ \overline{x} \in \{0,1\}^n$ & \\
\hline
\end{tabular}
\caption{Transformation between bipolar and binary representations.}\label{tab:xform}
\end{table}

Within the BP group, max-product BP fails to exploit the sparsity of the
measurement matrix, and CSBP assumes a sparse (not necessarily binary) signal.
Our NBP uses the same framework as CSBP, but with the correct fault  prior. As we show
shortly, using the correct prior results in a significant improvement in identifying the correct fault
pattern. LDLC  assumes a binary prior, but assigns faults and non-faults
equal probability, which degrades performance.

Within the linear programming group, linear programming and semidefinite programming
relax the binary fault prior into the continuous domain, returning 
fractional results, and sparsity of the fault pattern is not assumed or exploited.
Within the CS group, the binary nature
of the signal is not exploited.

\subsection{Experimental settings} \label{sec:exp:settings}
We consider an example with $m=50$ sensors, $n=100$ possible faults,
and linear measurements.
For simplicity, we assume that all faults are equiprobable, i.e.,
$p_s=p$ for all $s$.
The matrix $A$ is taken to be sparse with a fixed
percentage $q$ of non-zero entries, whose values are drawn
from a bipolar Bernoulli distribution (non-zero elements of $A$ are chosen randomly
and independently to be either $+1$ or $-1$).
Bipolar matrices often occur, for example, in fault identification of linear
analog circuits \cite{analog1,analog2} (Section \ref{analogc}).
We fix the measurement noise standard deviation to $\sigma = 1$.
 
%

We use word error rate (WER) to measure the fraction of problems on which we recover the
correct fault pattern exactly.
We also compare the methods using the standard precision/recall measure, where
precision is the number of true positives (correctly identified faults) divided by
the total number of identified faults (including the false positives), and recall
is the number of true positives divided by the total number of true positive and false negatives.

\ifthenelse{\boolean{TwoColumn}}{
  \input{figures23.inc} 
}{}
\ifthenelse{\boolean{TwoColumn}}{}{
  \input{figures23.inc} 
}
\subsection{Discussion of results} \label{sec:exp:discuss}
We show two sets of results, varying the sparsity $p$ of the fault patterns (rarity of faults) in
Fig.~\ref{fig:error} and the sparsity $q$ of fault signatures (the matrix $A$) in
Fig.~\ref{fig:sparsity}.  In both cases, we compare all ten algorithms listed in
Table~\ref{tab:methods} with and without local optimization heuristics.
At each sparsity level, the plots show performance averaged over 1000 experiments; vertical error bars indicate the corresponding confidence intervals.
Unless otherwise stated, we use default parameters for fault  sparsity $p=0.12$ and fault signature sparsity $q=0.2$.

We  evaluate the effect of fault pattern sparsity on solution quality.
Fig. \ref{fig:error} shows the performance of each algorithm, where the
sparsity level $q$ of $A$ is fixed to be 0.2, and the  fault prior $p$ varies between 3\% and 15\%.
When $p$ increases, the problem becomes harder, because there are an increasing
number of \emph{a priori} likely fault combinations.  For example, with a prior of $1/n$ we have on average
only one fault, with $n$ possible locations; when the prior is $2/n$ there are on average two faults
and $n(n-1)/2$ possible locations, and so on. Increasing the prior fault probability shows a
clear separation between the performance of the different algorithms, in which NBP
outperforms IP in all cases (both with and without local optimization).

Fig.~\ref{fig:sparsity} corresponds
to fault probability $0.12$ across a range of fault signature (the matrix $A$)
sparsities. It is clear that when the fault signature matrix is less sparse, the problem becomes easier, and the performance of all the algorithms improves. Again, NBP has the lowest WER in all tested scenarios (without
local optimization).
When using local optimization, IP performs better on dense matrices (when 45\% of the matrix entries are non-zeros).

We  note the effect of the local optimization heuristics on the quality of the solutions.
In both Figs. \ref{fig:error},\ref{fig:sparsity} the optimization heuristics have a positive effect, and this
effect is particularly powerful when the probability of a fault is low.

Regarding running time, shown in Fig. \ref{fig:runtimeprec}(a),(b), NBP  is
comparable to other BP algorithms and requires a few
seconds. However, linear programming algorithms such as IP and SDP are an order of
magnitude faster.
This suggests that when accuracy is the priority, NBP is better, but when
a fast approximation is sufficient, IP should be used instead. A similar conclusion was offered in \cite{CSBP2009}. 

Fig. \ref{fig:runtimeprec}(c) shows a standard precision/recall curve, averaged over 1000 experiments. NBP clearly dominates other algorithms in all ranges.
The closest performance to NBP is obtained using both IP\ and SDP.

\begin{figure}[t]
  \begin{center}
    \subfloat[Running time]{
      \label{fig:mo_a}
      \includegraphics[scale=0.22,clip]{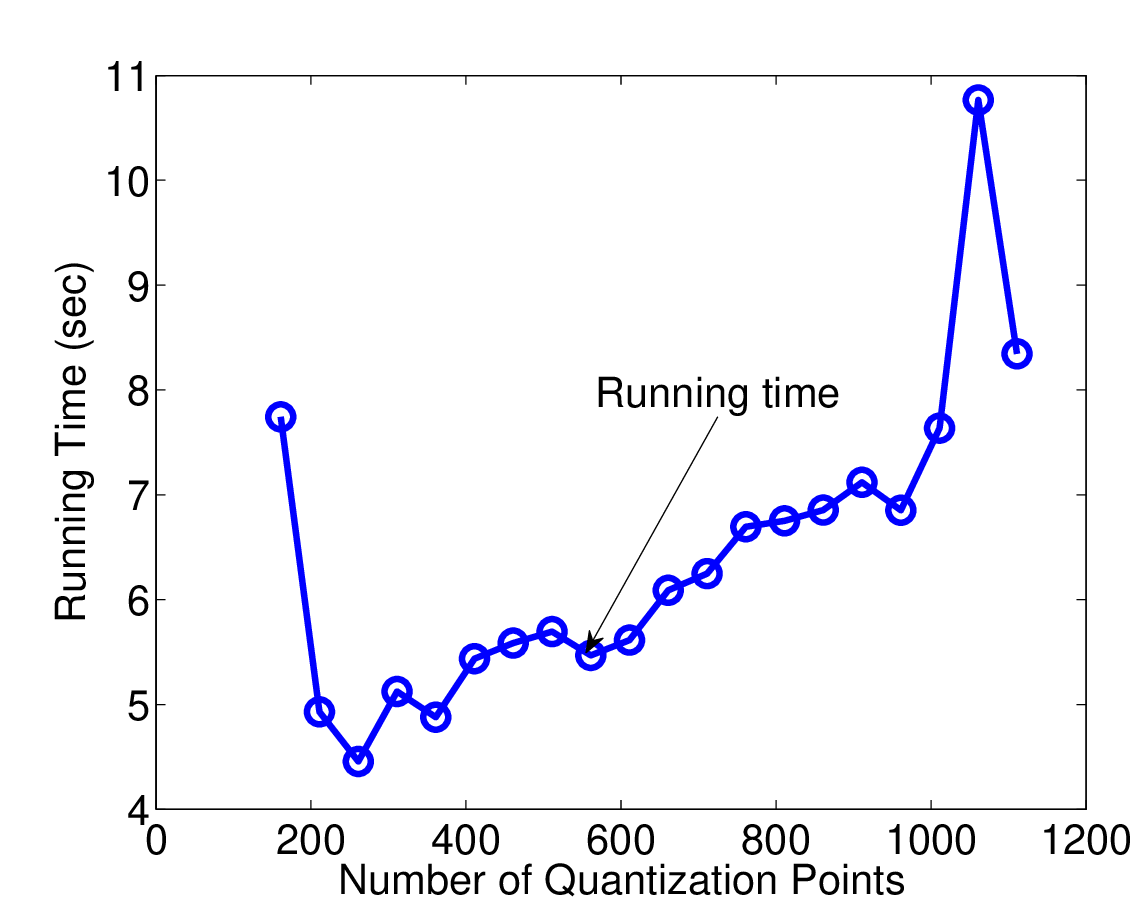}
    }
   \subfloat[Word error rate]{
     \includegraphics[clip,scale=0.22]{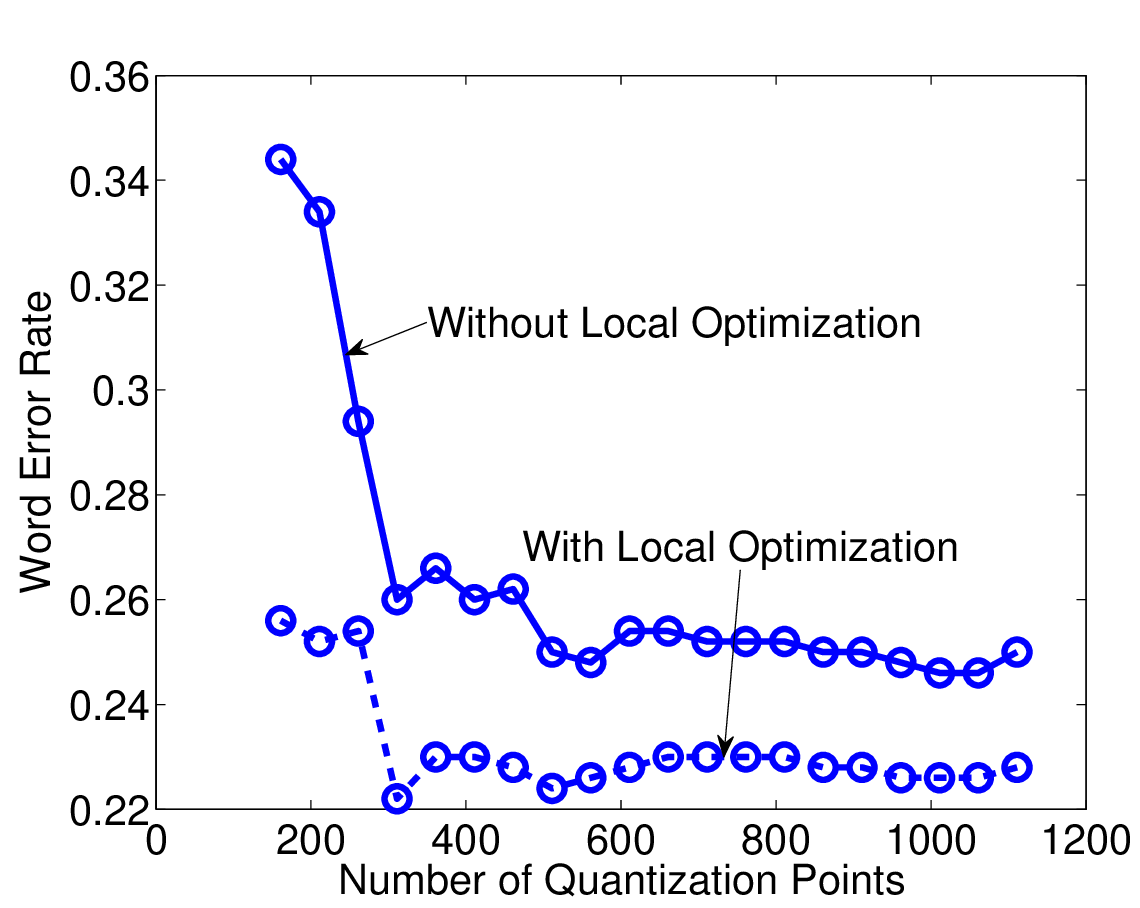}
     \label{fig:mo_b}
   }\\
      \subfloat[Noise WER]{
 \includegraphics[clip,scale=0.22]{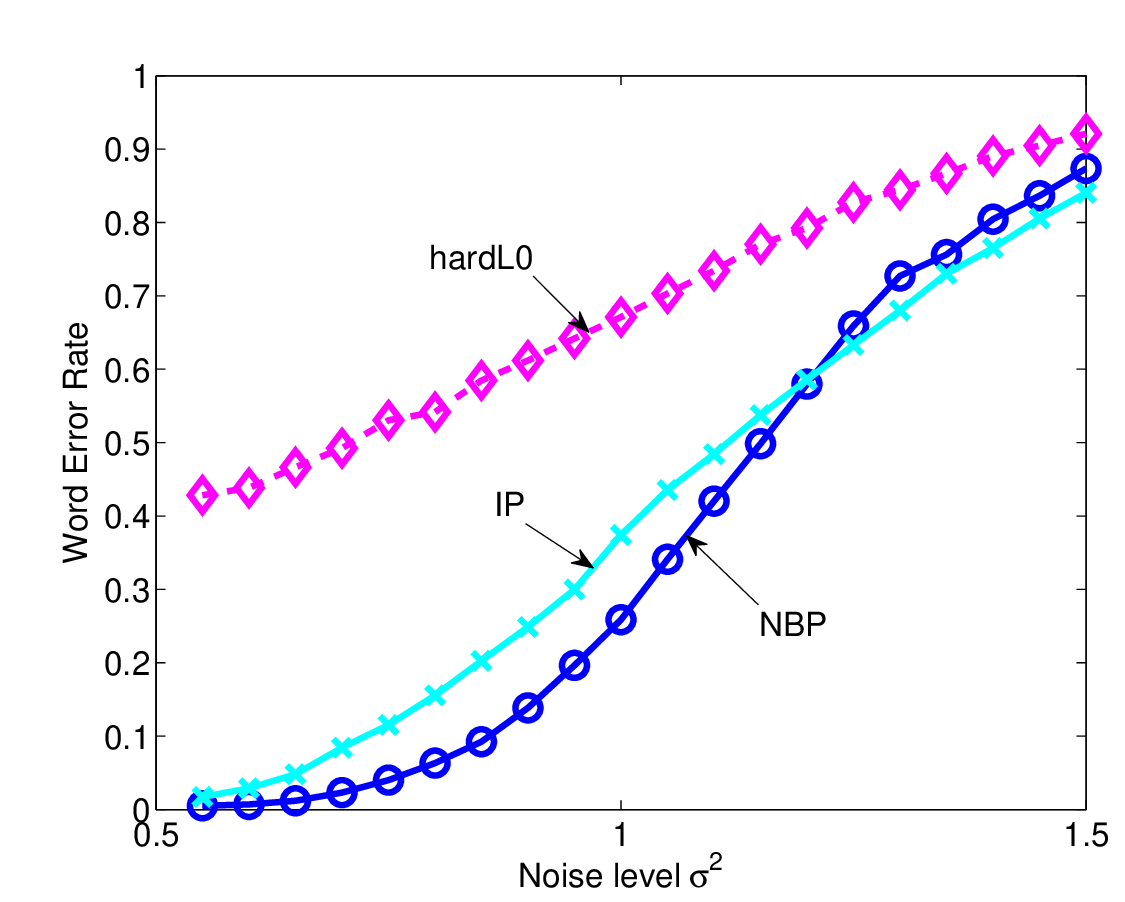}
     \label{fig:ns_c}
     }
      \subfloat[Robust WER]{
 \includegraphics[clip,scale=0.22]{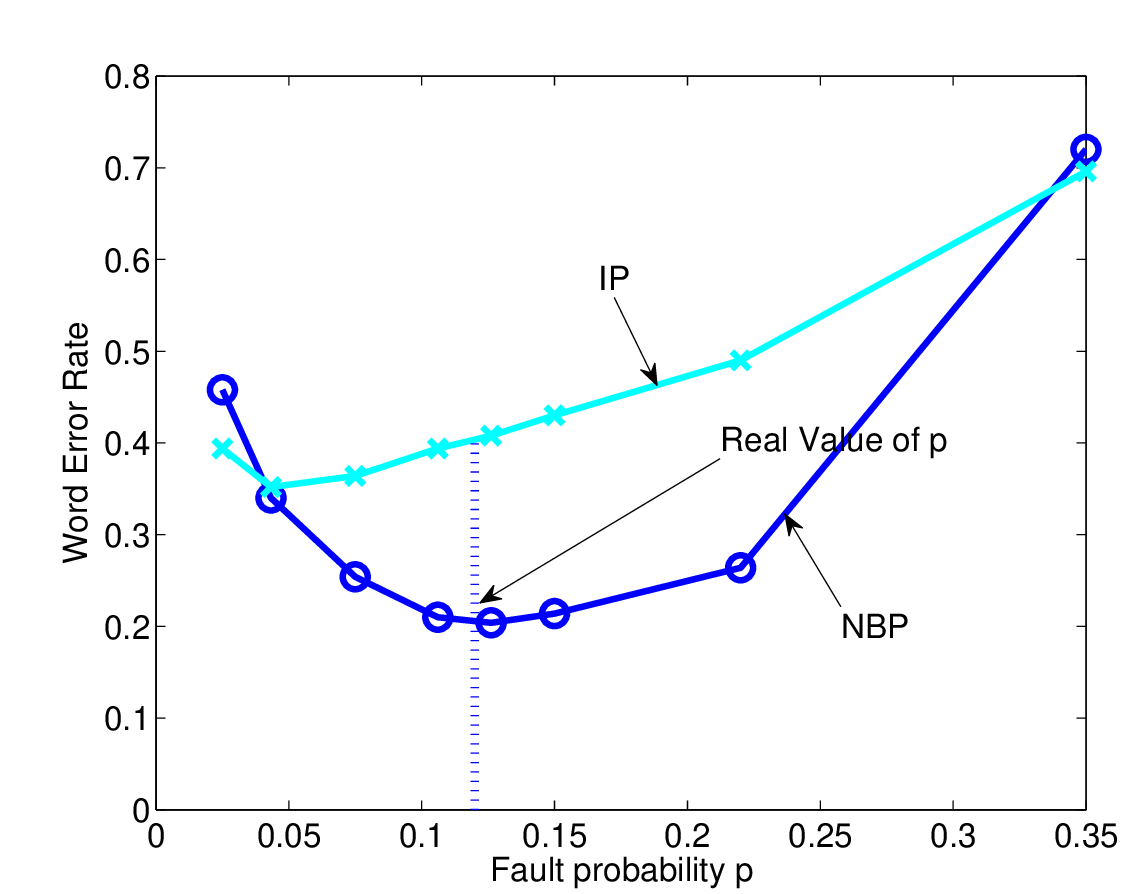}
     \label{fig:ns_e}
     }
   \caption{Top row: The effect of quantization on the (a) running time of NBP,
and (b) accuracy of NBP.  Increasing quantization slows NBP, but below a certain
point can significantly degrade performance.
Bottom row: the effect of (c) changing observation noise variance $\sigma^2$ on
WER, and (d) using the wrong fault prior $p$ on WER (the true fault probability was $p = 0.12$).}
\label{fig:changes}
\end{center}
\end{figure}

Next we evaluate how changes in the algorithm setup affect performance. Fig. \ref{fig:changes}(a),(b) investigates how the number of quantization points affect both performance and accuracy of the solution.
There is clearly a tradeoff between quantization level and efficiency. When fewer quantization points are used, the algorithm runs faster but with reduced accuracy.
Fig. \ref{fig:changes}(c) investigates the effect of the noise level $\sigma^2$
on performance.
As expected, as the noise increases the problem becomes more difficult and the error in identifying the correct fault patterns is larger.

\begin{figure}[tb] \centering
\includegraphics[width=.375\textwidth]{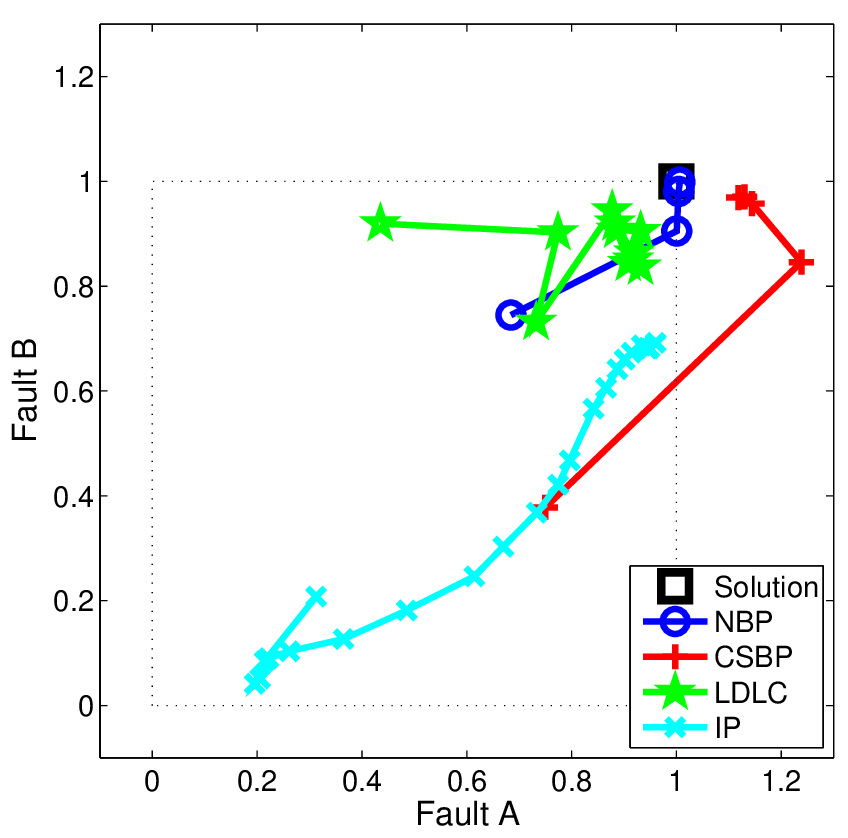}
\caption{Convergence of the soft decision values for two faulty bits using different algorithms.}
\label{fig:3D}
\end{figure}

Another issue of interest is that NBP requires us to know the statistics of the fault
patterns, and the performance of NBP may deteriorate if the estimate of the
fault sparsity $p$ is wrong. To demonstrate that the deterioration is not severe,
Fig \ref{fig:changes}(d) shows the effect of an incorrect estimate of
$p$ on the algorithms' performance. The true fault sparsity was taken to
be $p=0.12$, but a false value of $p$ is reported to the algorithm
($x$-axis).  NBP remained more accurate than the other methods even with moderate
model mismatch.

\ Finally, we illustrate the convergence of NBP, CSBP, LDLC, and IP in Fig. \ref{fig:3D},
where $m=10, n=15$, and two faults occurred.
Each axis indicates the belief or soft decision around one of the two (correct) faults.
Note that all algorithms converge to solutions greater than 0.5, and will thus
round to the correct solution in post-processing.
The NBP algorithm converges in two iterations to the correct solution
while IP  requires more iterations for converging to an approximate solution,
although we should be cautious in giving this conclusion too much weight, since
the computational cost per iteration of the NBP algorithm is higher.
NBP converges accurately, because the
prior distribution encourages it to converge to zero or one.
CSBP converges to some positive but non-integral value, while LDLC also
encourages binary values and is therefore the closest after NBP.

Overall, NBP has better performance for fault
identification in most of the cases tested. The best competitor is IP, which
performed better when the fault signature matrix was dense and
local optimization heuristics were applied. The main benefits of NBP are
a) it is a distributed algorithm that can scale to large problems (see for
example \cite{LDLC_Sommer}), whereas other algorithms are centralized;
b)\ it performs better when the fault signature matrix is sparse or the
fault vector is dense; and
c)\ local optimization heuristics, which require more computation, are less
important than for other algorithms.
Finally, IP is subject to numerical errors (see discussion of numerical
errors in \cite{ISIT09-3}), which we encountered when running on larger
problem sizes (e.g., $m=400, n=200$).
In contrast, we did not encounter numerical errors when using NBP.

The drawbacks of NBP are
a)\ it is slower than IP; and
b) it has several configurable parameters that need to be fine-tuned,
including the quantization level, quantization bounds, and the variance
of the local potential approximation, \eqref{gauss_approx}.

\section{Information Theoretic Characterization}
\label{sec:inf_theory}
Why does NBP perform so well relative to state-of-the-art
algorithms? A partial answer can be obtained by considering recent information
theoretic results in the related domains of multi-user
detection~\cite{GuoVerdu2005,GuoWang2008} and CS~\cite{GuoBaronShamai2009,RFG2009,limits}.
Consider the signal and measurement models as before, where $y = A x + v$,
$x_s$ is i.i.d.\ Bernoulli with $\Pr(x_s=1)=p$, and $v$ is i.i.d.\ zero mean unit
norm Gaussian. Suppose also that the measurement matrix $A$ is assumed to be i.i.d.\
where the probability of non-zero entries is $q$, and that non-zero matrix entries
follow some unit norm distribution. We note in passing that the non-zero probability
$q$ must scale to zero as $m$ and $n$ grow; see details in Guo and Wang~\cite{GuoWang2008}.

Following the conventions of Guo et al.~\cite{GuoBaronShamai2009,GuoWang2008,GuoTanaka2009},
the signal to noise ratio (SNR) $\gamma$ can be computed as,
\[
\gamma = mq.
\]
We also define a distortion metric $D$ that evaluates the approximation error between $x$
and $\widehat{x}$ by averaging over per-element distortions,
\[
D(x,\widehat{x}) = \frac{1}{n} \sum_{s=1}^n d(x_s,\widehat{x_s}),
\]
where $d(\cdot,\cdot)$ is a distortion metric such as square error or Hamming
distortion. For this problem where a sparse measurement matrix $A$
is used, Guo and Wang~\cite{GuoWang2008} provided the fundamental information theoretical
characterization of the optimal performance that can be achieved,
namely the minimal $D$ that can be achieved as $m$ and $n$ scale to infinity with
fixed aspect ratio $\delta$, i.e., $\lim_{n\rightarrow\infty}\frac{m}{n}=\delta$.

There are several noteworthy aspects in the analysis by Guo and Wang.
First, Theorem~1~\cite{GuoWang2008} proves that the problem behaves as if
each individual input element $x_s$ were estimated individually from a corrupted version
$z_s$, with $z_s=x_s+w_s$ where $w_s$ is Gaussian noise.
That is, the vector estimation problem is related to an estimation problem defined
over scalar Gaussian channels.
Second, Theorem~2~\cite{GuoWang2008} demonstrates that the
amount of scalar Gaussian noise in the random variable $W_s$ can be computed by finding
the fixed point for $\eta$ of the following expression:
\begin{equation} \label{eq:eta}
\eta^{-1} = 1 + \frac{\gamma}{\delta} \mbox{mmse}(p,\eta \gamma),
\end{equation}
where $\eta\in(0,1)$ is the degradation in SNR, i.e., the SNR of the scalar channel
is $\eta\gamma$ instead of $\gamma$, and $\mbox{mmse}(p,\eta \gamma)$ is the minimum
mean square error (MMSE) for estimating the random variable $X_s\sim\mbox{Ber}(p)$
from the output $Z_s$ of the dual scalar channel whose SNR is $\eta\gamma$.
That is, the scalar Gaussian channels are degraded.
Combining these insights, the fundamental limiting performance can be computed for any
distortion metric $d(\cdot,\cdot)$ by examining the output of the degraded scalar channel with
SNR $\eta\gamma$. Finally, Guo and Wang demonstrate that in the large system limit (when $m$ and $n$ scale to infinity) with
fixed aspect ratio $\delta$, the posteriors estimated by BP converge in distribution
to the true posteriors. Consequently, it should be no surprise that numerical results in
Section~\ref{sec:exp_results} indicate that NBP outperforms other techniques for fault
identification in practice.

\begin{figure}\centering
\includegraphics[width=.75\columnwidth]{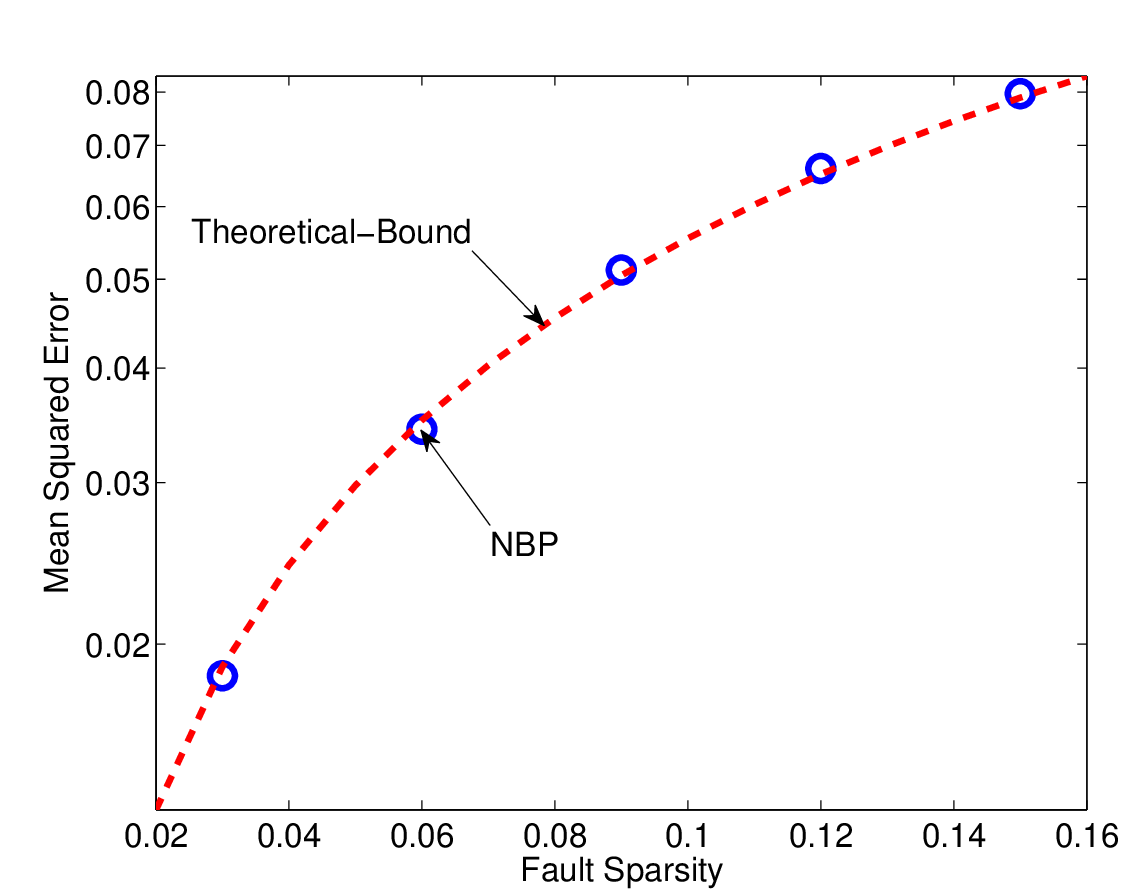}
\caption{Mean squared error results for NBP as a function of fault sparsity,
compared to the theoretical lower bound on MSE performance.  NBP provides nearly
optimal estimates even on modestly sized problems.
}
\label{fig:MSE}
\end{figure}

To verify that our NBP algorithm indeed offers MSE performance that is
comparable to the
information theoretic lower bound (\ref{eq:eta}), we simulated a
setting similar to that
of Section~\ref{sec:exp_results} where $m=250$, $n=500$, $q=0.1$,
$\sigma=4$, and
varying $p$ from $0.03$ to $0.15$. Figure~\ref{fig:MSE} illustrates
that the mean square error achieved by NBP is almost indistinguishable from 
the MMSE bound \eqref{eq:eta}. 

Although the fundamental performance of fault identification is well understood
when $A$ is sparse and i.i.d., when $A$ is not sparse the analysis becomes
more involved. In the non-sparse case, replica method analyses have been used to
derive information theoretic characterizations in a less rigorous fashion; 
see Guo and Verd\'u~\cite{GuoVerdu2005} or Guo and Tanaka~\cite{GuoTanaka2009}.

\vspace{-1mm}
\section{Discussion}
\label{sec:conclusion}

In this paper, we  propose a novel approach for solving the fault identification
problem using non-parametric belief propagation. By exploiting prior
information about faults, we are able to improve significantly our ability to
correctly identify fault patterns.  Our main improvement arises from accounting for
both the binary nature and prior probability of the faults.

Our approach elects to quantize a continuous relaxation of the true discrete model
before solving it using belief propagation (BP).  Although this step introduces
additional approximations, it enables us to exploit the fact that each likelihood
function $f_i$ is a function of a linear combination of variables $x_s$.
This makes the algorithm efficient even when $q$ is not small, so that the $f_i$
are defined on a large number of variables.

A possible extension is the inclusion of non-i.i.d.\ noise.  In principle,
non-i.i.d.\ noise can be dealt with by augmenting the model in Fig.~\ref{fig:factorgraph}(a) with
a graph structure over measurements $y$ that reflects the inverse covariance matrix of the noise.
We expect our approach to continue to do well for relatively sparse inverse
covariance matrices;  an exact characterization is left for future study.

Another extension would be to consider non-binary faults. As long as the fault
pattern is i.i.d., the recent information theoretic results of Guo et
al.~\cite{GuoBaronShamai2009,GuoWang2008,GuoTanaka2009}
indicate that BP should continue to perform well.

As a final note, we mention that our proposed algorithm is distributed, since the underlying BP algorithm
can be distributed over multiple nodes, and works well when the matrix
$A$ is sparse. In a network where communication is costly,
our algorithm offers the additional advantage of requiring less communication.

\section*{Acknowledgements}
Danny Bickson was partially supported by grants ARO MURI W911NF0710287, ARO MURI W911NF0810242, NSF Mundo IIS-0803333 and NSF Nets-NBD  CNS-0721591.
Parts of this work were performed while Danny Bickson was a research staff member at IBM\ Haifa Labs. Dror Baron thanks the Department of Electrical Engineering at the
Technion for generous hospitality while parts of the work were
performed, and in particular the support of Tsachy Weissman.
Danny Dolev is Incumbent of the Berthold Badler Chair
in Computer Science, and was supported in part by the Israel Science Foundation (ISF) Grant  0397373.

{\small
\bibliographystyle{IEEEtran}
\bibliography{IEEEabrv,used}
}

\end{document}

%% file: table1.inc
\begin{table*}[t] \centering
\begin{tabular}{|l|l|l|l|c|c|}\hline
Group
& Algorithm & Abbreviation & Prior on x  \\\hline\hline
BP
& \textbf{NBP Solver} (current work)  & NBP & binary and sparse \\
 & \textbf{Max-product BP}  (current work)& Max-prod & binary and sparse \\
&  Compressed sensing Belief Propagation \cite{CSBP2009}& CSBP & sparse \\ 
& Low density lattice decoder \cite{LDLC_Sommer}& LDLC & binary \\\hline
LP
& Interior point (Newton method) \cite{FaultDet}& IP &$x\in[0,1]$  \\
&Semidefinite programming\cite{TR:01,ANJM:02}& SDP & $x\in[0,1]$  \\\hline
CS
& Iterative signal recovery \cite{CoSaMP}& CoSaMP & sparse \\
& Gradient Projection for Sparse Reconstruction \cite{GPSR}& GPSR & sparse \\
& Iterative hard thresholding \cite{hardIO}& hardIO & sparse  \\\hline
Other & All zero hypothesis & Null & $x$ is constant  \\ \hline
\end{tabular}
\caption{Algorithms used for comparison, grouped into general categories: belief propagation methods (BP),
linear programming (LP), and compressed sensing (CS).}
\label{tab:methods}
\end{table*}

%% file: figures23.inc
\begin{figure*}[ht!]
\vspace{-6mm}
  \begin{center}
    \subfloat[WER without local optimization]{
      \label{fig:err_a}
      \includegraphics[scale=0.34,clip]{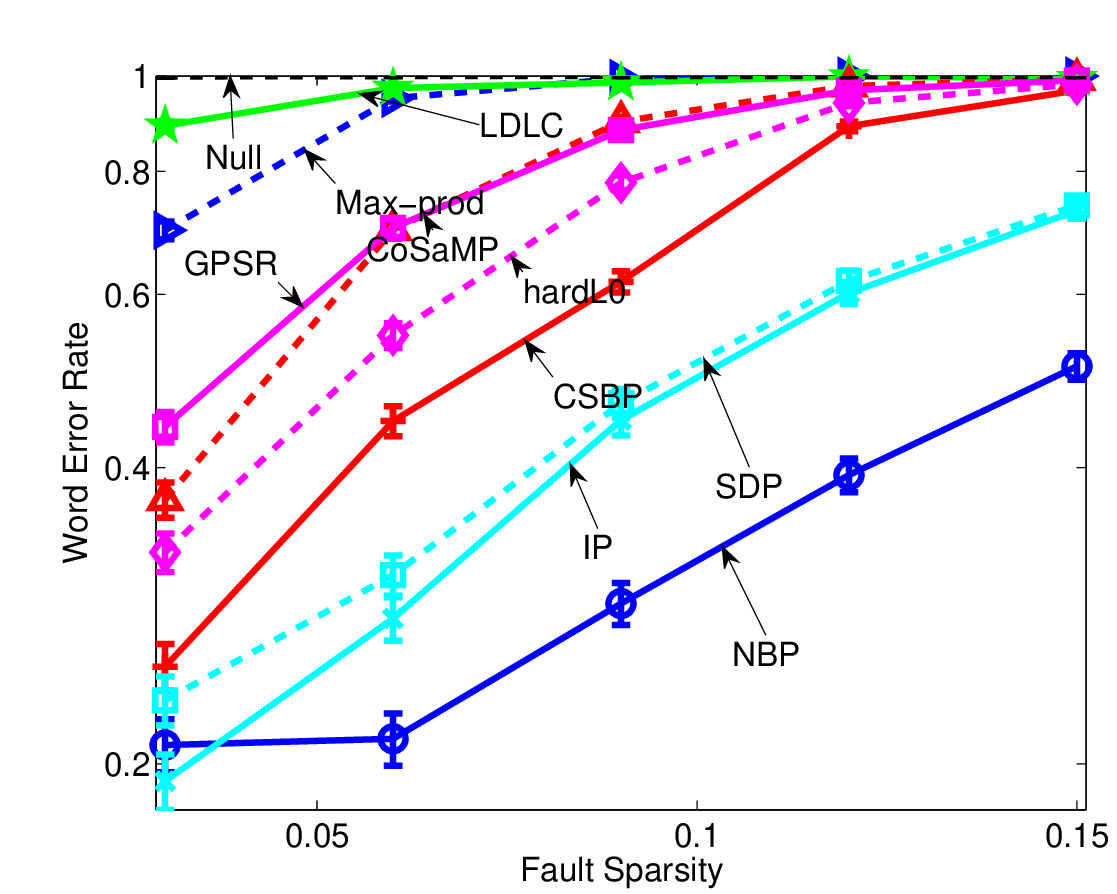}
    }
   \subfloat[Legend]{
      \label{fig:err_legend}
      \includegraphics[scale=0.44,clip,bb=400 150 536 410]{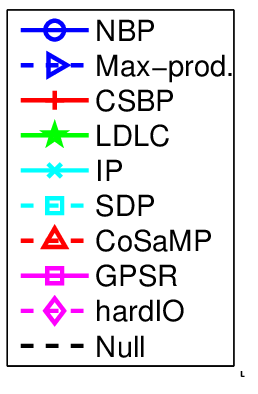}
    }
   \subfloat[WER with local optimization]{
     \includegraphics[clip,scale=0.34]{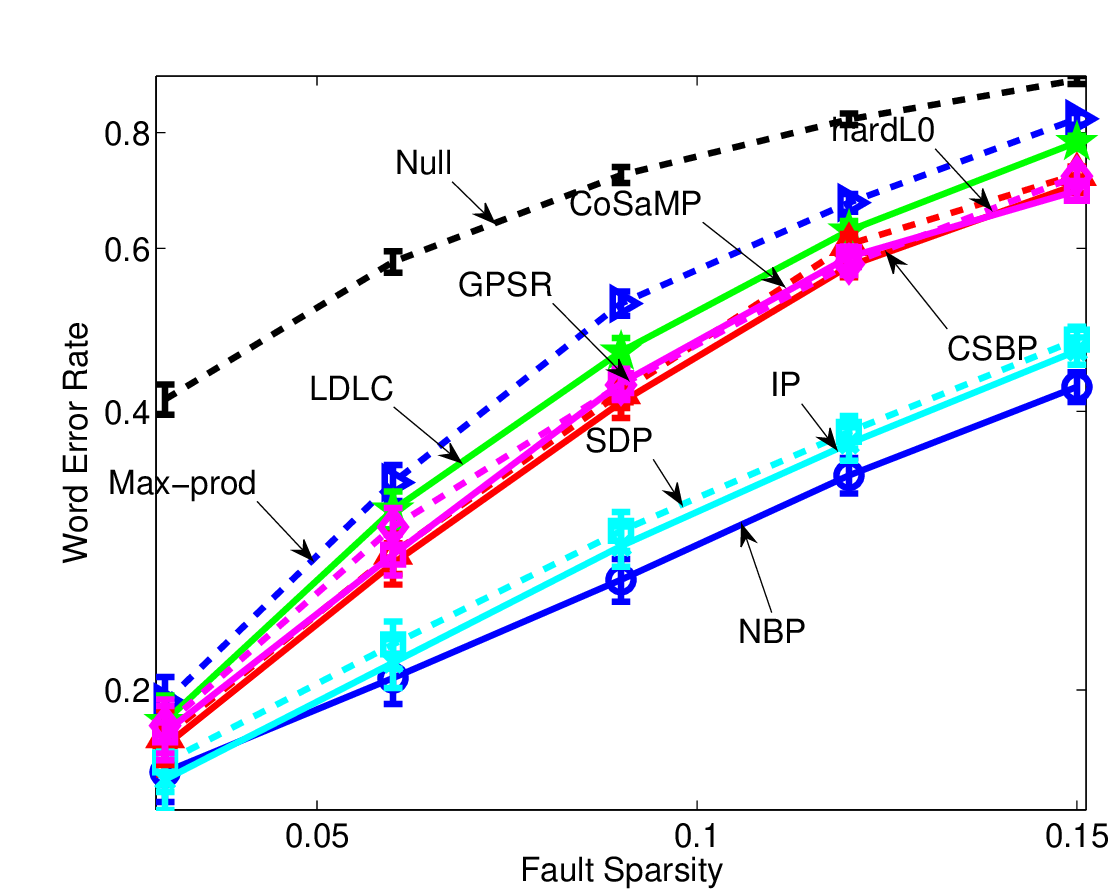}
     \label{fig:err_b}
   }\\
   \caption{The effect of changing the probability $p$ of faults on the reconstruction success of the different methods. }
\label{fig:error}
\end{center}
\end{figure*}
\begin{figure*}[ht!]
\vspace{-10mm}
  \begin{center}
    \subfloat[WER without local optimization]{
      \label{fig:spr_a}
      \includegraphics[scale=0.34,clip]{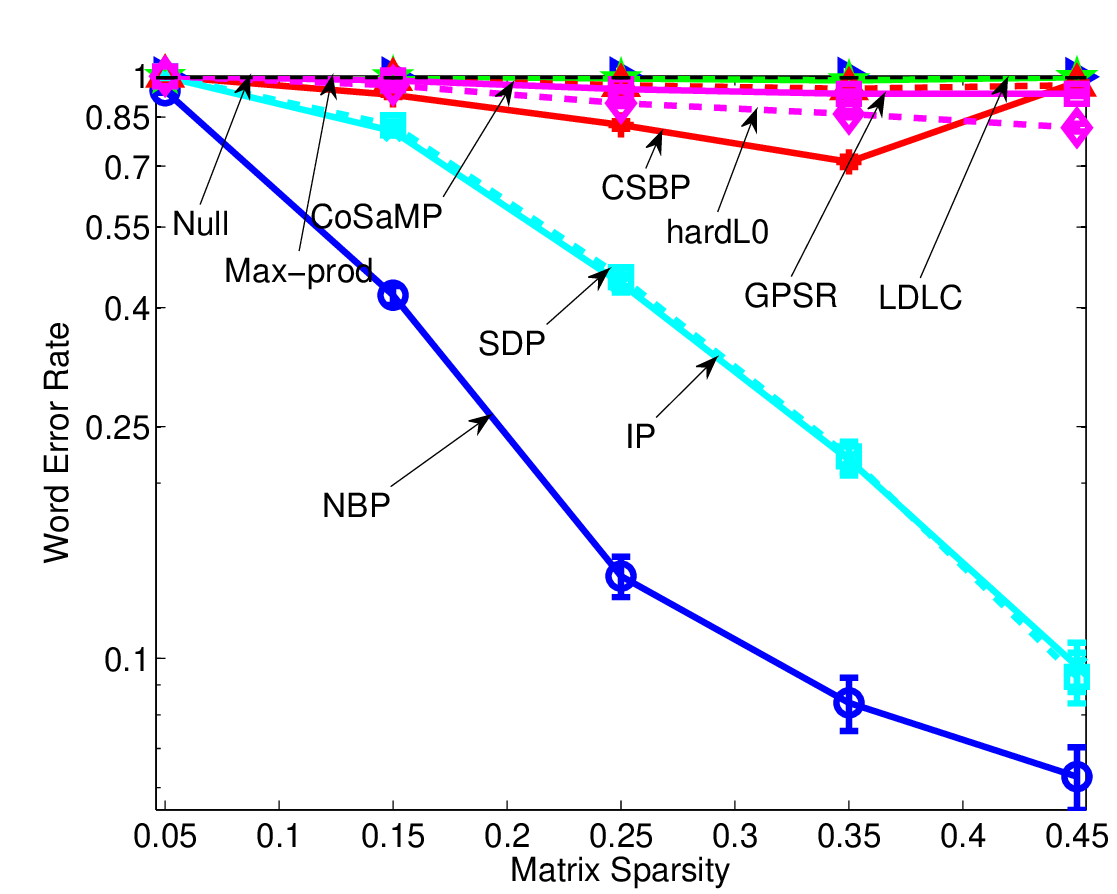}
    }
     \subfloat[Legend]{
      \label{fig:spr_legend}
      \includegraphics[scale=0.44,clip,bb=400 150 536 410]{fig/exLegend.eps}
    }
   \subfloat[WER with local optimization]{
     \includegraphics[clip,scale=0.34]{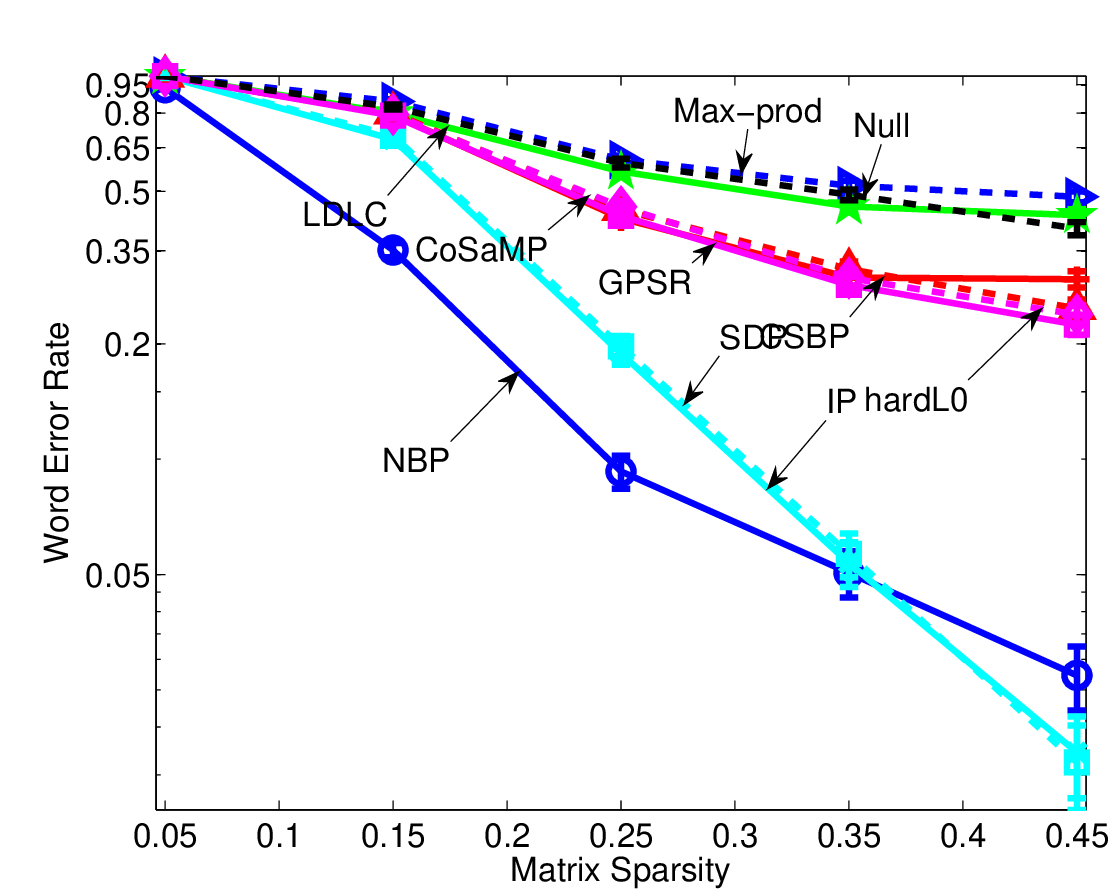}
     \label{fig:spr_b}
   }\\
   \caption{The effect of changing the fault signature matrix sparsity $q$ on the reconstruction success of the different methods. }
\label{fig:sparsity}
\end{center}
\end{figure*}

\begin{figure*}[ht!]
\vspace{-8mm}
  \subfloat[Running time vs. fault sparsity]{
 \includegraphics[clip,scale=0.30]{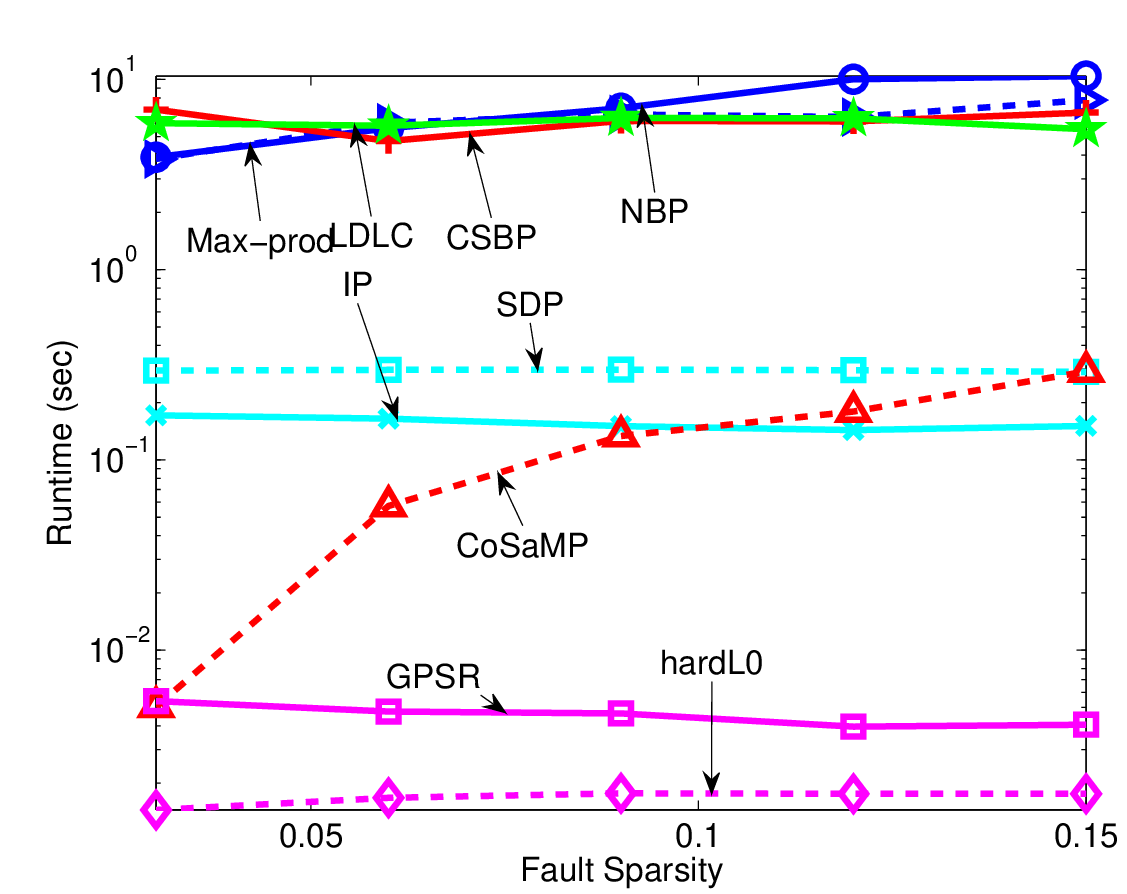}
     }
    \subfloat[Running time vs. matrix sparsity]{
 \includegraphics[clip,scale=0.30]{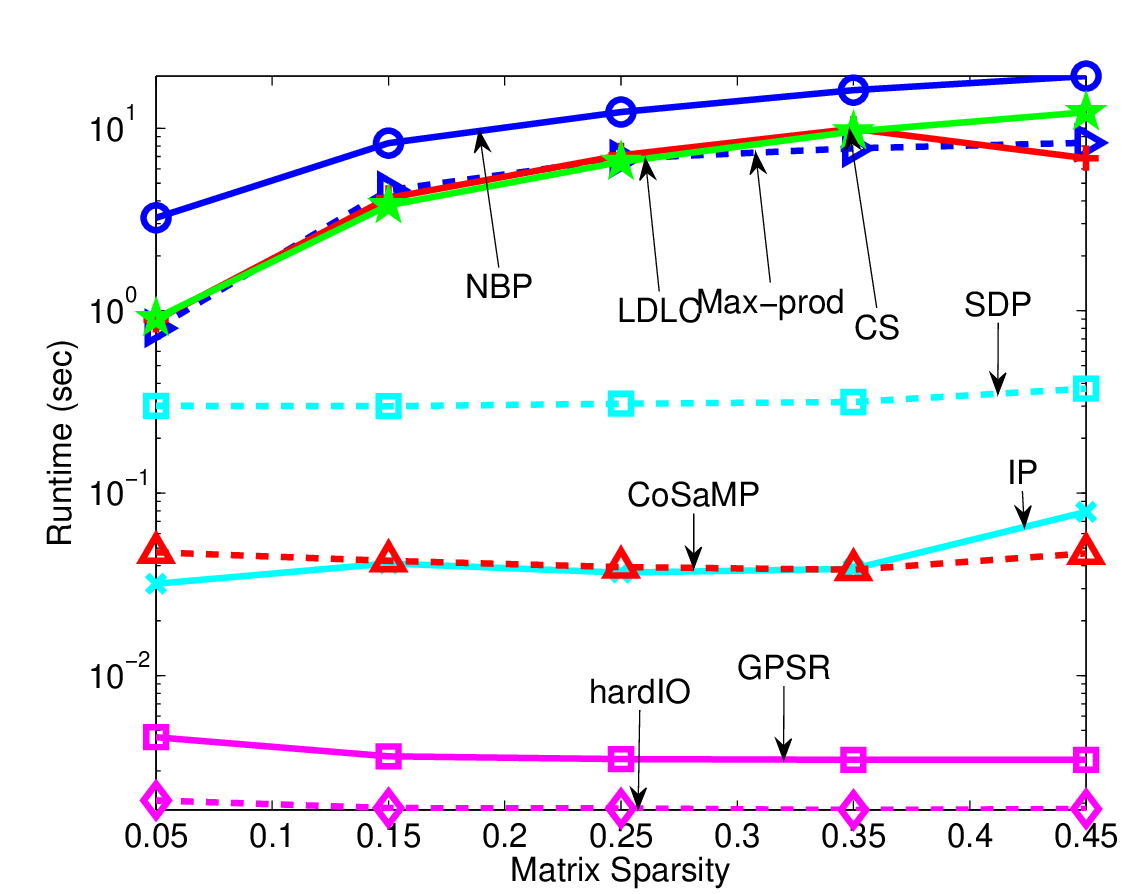}
   }
   \subfloat[Precision/Recall curve]{
 \includegraphics[clip,scale=0.30]{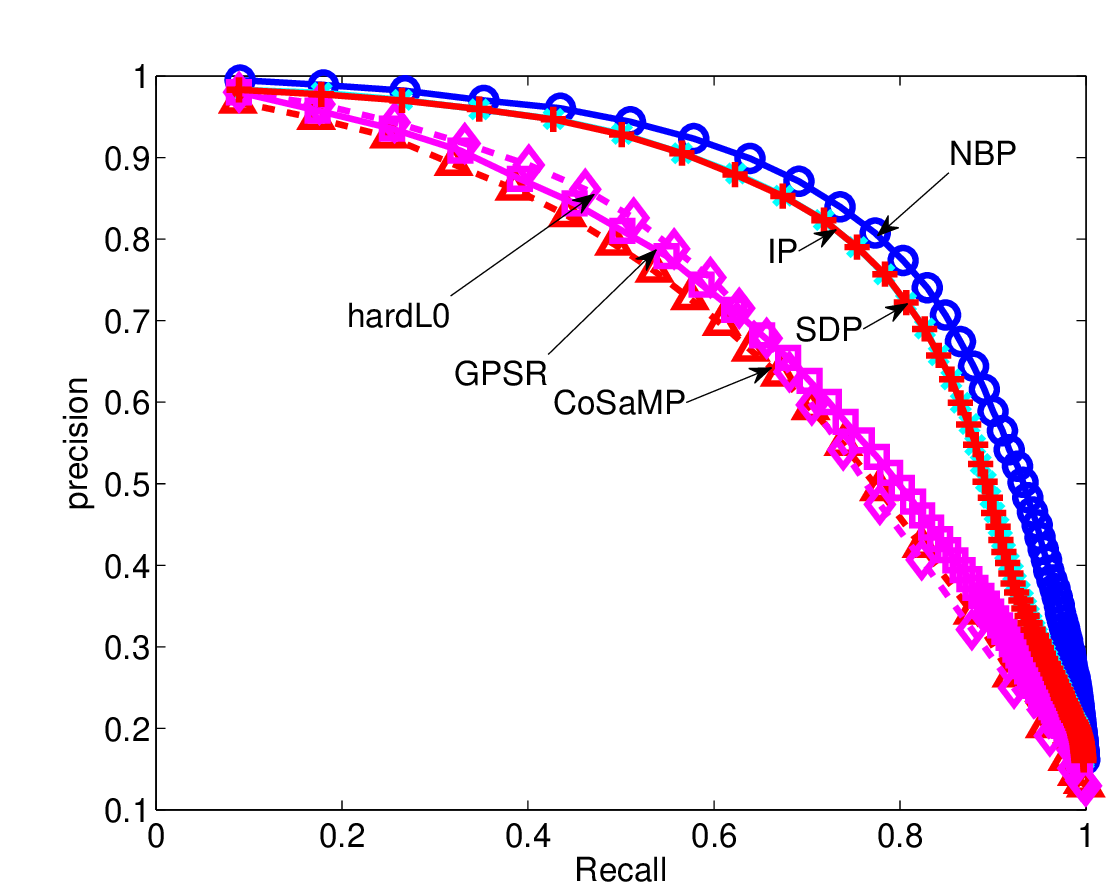}
   }
   \caption{(a),(b) Running times of the different algorithms across experimental conditions; (c) precision/recall curve for each method, at $p=0.12$ and $q=0.2$, as the fault decision threshold is varied.}
\label{fig:runtimeprec}
\end{figure*}